\documentclass[11pt]{article} 
\usepackage{amsfonts,amscd,amsmath,graphicx}

\def\NP#1#2{Nucl.Phys. B#1 (#2)} 
\def\PL#1#2{Phys.Lett. B#1 (#2)}
 
\def\MPL#1#2{Mod.Phys.Lett.A#1 (#2)} 
 
\def\PR#1#2{Phys.Rev. D#1 (#2)}

\def\HP#1#2{JHEP #1 (#2)} 
\def\ap{ \alpha^{\prime}} 
\def\ai{\pi\alpha^{\prime}}
\def\2p{2\pi\alpha^{\prime}}
\def\pd{\partial}

\def\tr{\text{tr}}
\newcommand{\ep}{\text e}
\newcommand{\oh}{\frac{1}{2}}
\title{More About Partition Function of Open Bosonic String in Background Fields 
and String Theory Effective Action}
\author{Oleg Andreev\thanks{e-mail:  andreev@physik.hu-berlin.de}
\thanks{Permanent address: Landau Institute for Theoretical Physics, Moscow, Russia}
\\ \\
Humboldt--Universit\"at zu Berlin, Institut f\"ur Physik\\
Invalidenstra\ss e 110, D-10115 Berlin, Germany}

\date{}
\begin{document} 
 
\maketitle 
\begin{abstract} 
We refine the relation between the renormalized partition function of the open bosonic string in 
background fields and string theory effective action. In the process, we get some leading 
derivative corrections to the Born-Infeld action which include all powers of the gauge field. 
\end{abstract}

\vspace{-12cm}
\begin{flushright}
hep-th/0104061      \\
HU Berlin-EP-01/15
\end{flushright}
\vspace{11cm}


\vspace{.1cm}
Open strings in background fields were intensively studied in the eighties  as their partition 
functions provide the low energy effective actions of string modes, for example Dirac-Born-Infeld action \cite{FT,AC,P}, that include all 
powers of fields. Later it was also realized that it is a rather useful tool for D-brane physics \cite{P1}. Recently, there 
has been a resurgence of interest in this subject related with the study of tachyon 
condensation (see, e.g., \cite{t} and references therein). The purpose of this paper is to refine the relation between 
the renormalized bosonic open string partition function and  the string theory effective action that provides a practical method for 
getting  derivative corrections including all powers of fields to the Born-Infeld action.

In fact, the idea that the effective action of string modes is provided by the renormalized partition function has a long 
history. It has been already suggested in \cite{F,Tt} that the partition function is a natural object to go off-shell.  Another 
famous example emerged in the context of the Witten's background-independent open string theory \cite{W}. 

To further clarify the relation let us first give some useful information about the string 
theory effective action. Since it should be 
consistent with the string S-matrix near the perturbative vacuum, we have 
\begin{equation}\label{i}
S_{\text{eff}}\vert_{\text{pert.vac.}}=S_{\text{S-mat.}}
\quad,
\end{equation} 
where $S_{\text{S-mat.}}$ is an effective action of string modes reconstructed from on-shell string amplitudes within the so-called S-matrix approach (see, e.g., \cite{Sh}). Although 
Eq.\eqref{i} is very natural to require, it becomes somewhat puzzling as soon as we define the effective action via the 
partition function. Indeed, $S_{\text{S-mat.}}$ is a well-defined object via the string S-matrix while the partition function, 
as a 2d Field Theory object, may suffer from 
divergences. So one has to use the whole machinery of the 
renormalization 
group (RG) to take the situation under control. From this, it is unsurprising that the left hand 
side computed in a given renormalization scheme may not coincide with the right hand side. It is 
usually believed \cite{Tl} that after a proper redefinition of fields in $S_{\text{S-mat.}}$ 
which is of course harmless for the S-matrix both sides of Eq.\eqref{i} coincide. Actually, it is partially true. Indeed, 
as long as one only deals with logarithmic singularities as in the case of superstring it works 
perfectly. But it fails as soon as  power like singularities enter the game. This is exactly 
what happens in the case of bosonic string. In fact, this puzzle
was first observed in \cite{AT} (see also \cite{AT3}) where it was shown 
that the renormalized partition function in general does not reproduce string amplitudes for 
the gauge field \footnote{Strictly speaking, the situation 
is more involved as the effective action can be defined via the partition function in a 
non-trivial way. However, one does not run into trouble as we will see later.}. Some 
possible resolutions of the problem are:

(i) to treat power like singularities in the same way as it is done in string theory 
namely, to use analytic continuation in external momenta and to fix the 
M\" obius gauge. However, it turns out very difficult to compute contributions to the 
partition function which include {\it all powers} of fields in such a way. 

(ii) to try more convenient schemes in order to approximate the string recipe 
as much as one can do. But it is a very tedious way to compute the contributions of 
interest \footnote{In fact, a scheme that reproduces the leading momentum dependence of the 
gauge field four point amplitudes has been recently proposed in \cite{A2, Tt2} but there is no 
guarantee that it reproduces the whole amplitudes.}. 

Instead of running into such difficulties, let us propose another resolution. In fact, this puzzle means that a space of renormalization schemes is larger than a space 
of field redefinitions. So, to take the situation under control we have somehow to impose constraints 
on a space of renormalization schemes. A way to do so is to set the renormalization 
conditions within the 2d Field Theory. It is useful to keep in mind Eq.\eqref{i} which in principle 
suggests to take some structures from $S_{\text{S-mat.}}$, i.e. from the S-matrix approach \cite{Sh}, 
as the renormalization conditions.  


Before demonstrating how our proposal works, we will make a detour and recall some basic points of the bosonic sigma model 
that are relevant for what follows (see e.g.,\cite{Tt2,T} and references therein). The world-sheet action is given by 
\begin{equation}\label{ac}
S=\frac{1}{4\ai }\int_{D}d^2z\,
\delta_{ij}\pd_aX^i\pd^aX^j\,
+\oint _{\pd D} d\varphi\, \text{T}(X)\,
-i\oint_{\pd D}ds\,A_i(X)\dot X^i
\quad,
\end{equation}
where $D$ means the unit disk in the complex plane, $\pd D$ - its boundary. $T,\,A_i$ are the tachyon and 
gauge fields, respectively. $X^i$ map the world-sheet to the brane and $i,j=1,\dots,p+1$. The world-sheet indices 
are denoted by $a,b$.

A natural object to compute is the $\sigma$-model partition function
\begin{equation}\label{pf}
\text{Z}[\text{T},A ]=\int {\cal D}X\,\ep^{-S}
\quad.
\end{equation}
In doing so,  one can follow the original approach  of \cite{FT} namely, first integrate over the internal points of the 
world-sheet to reduce the integral to the boundaries and then split the integration variable $X^i$ on the 
constant $x^i$ and non-constant parts $\xi^i$. As a result, we have
\begin{gather}
\text{Z}[\text{T},\text{A} ]=\int [dx]\,\langle\,\ep^{-S_b}\,\rangle
\quad,\\
S_b=\int _{0}^{2\pi} d\varphi\Bigl(\text{T}+\oh\pd_{ij}\text{T}\xi^i\xi^j-
\frac{i}{2}\text{F}_{ij}\dot\xi^j\xi^i
-\frac{i}{3}\pd_k\text{F}_{ij}\dot\xi^j\xi^i\xi^k
-\frac{i}{8}\pd_{kl}\text{F}_{ij}\dot\xi^j\xi^i\xi^k\xi^l+O\bigl(\ap\,{}^{\frac{3}{2}}\bigr)\Bigr)
\quad, \label{pf1}
\end{gather}
where $[dx]=\frac{d^{p+1}x}{(\2p)^{(p+1)/2}}$, $\text{F}_{ij}=\2p F_{ij}$ and 
$\pd_i=\sqrt{\2p}\frac{\pd}{\pd x^i}$. We have also rescaled the integration variable as 
$\xi\rightarrow \sqrt{\2p}\xi$, so the propagator is given by
\begin{equation}\label{pr}
\langle\,\xi^i(\varphi_1)\xi^j(\varphi_2)\rangle =\frac{1}{\pi}\delta^{ij}\sum_{n=1}^{\infty}
\frac{1}{n}\cos n\varphi_{12}
\quad\text{with}\quad
\varphi_{12}=\varphi_1-\varphi_2
\quad.
\end{equation}
It is always possible to include the $\pd^2 \text{T}\xi\xi$ and $\text{F}\dot\xi\xi$- terms into the propagator. 
However, 
for what follows we will include only the  second one. A simple algebra shows that in this case the propagator 
takes the form \cite{AC}
\begin{equation}\label{pr1}
\langle\,\xi^i(\varphi_1)\xi^j(\varphi_2)\rangle (\text{F})
=\frac{1}{2\pi}\Bigl(\gamma^{ij}
\sum_{n=1}^{\infty}\frac{1}{n}\ep^{in\varphi_{12}}
+\gamma^{ji}
\sum_{n=1}^{\infty}\frac{1}{n}\ep^{-in\varphi_{12}}\Bigr)
\quad,
\end{equation}
where $\gamma^{ij}=\bigl(\frac{1}{\delta+\text{F}}\bigr)^{ij}$.

What this means in practice is as follows. The expansion of the partition function in powers of $\ap$ is equivalent to 
its expansion in powers of derivative $\pd$ \footnote{This expansion assumes the use of F instead of $F$.}. However, 
this does not correspond to the loop expansion as it is clear 
from Eq.\eqref{pf1}. We will see  manifestation of this fact when we compute the leading 
derivative corrections. Thus we have 
\begin{equation}\label{pf2}
\text{Z}[\text{T,A} ]=\int [dx]\,\ep^{-2\pi\text{T}}Z_0 (\text{A})\Bigl(1+Z_2+Z_4+\dots\Bigr)
\quad,
\end{equation}
where $Z_2$ contains two derivatives, $Z_4$ - four, etc.

Now, we have all in our disposal to proceed. As a warmup, let us first show how the 
Born-Infeld (BI) action, i.e. the leading contribution $Z_0$ (determinant) in the expansion of the partition function, 
inevitably appears in our settings.

The partial differential equation 
\begin{equation}\label{eq}
\frac{\pd\ln Z_0}{\pd \,\text{F}_{ij}}=i\pi\langle\,\dot\xi^j(\varphi)\xi^i(\varphi)\,\rangle (\text{F})
\end{equation}
is simply solved by 
\begin{equation}\label{bi}
\ln Z_0=\tr\ln\gamma\sum_{n=1}^\infty 1
\quad.
\end{equation}
However, the above expression is pure formal as the sum is divergent. It is clear that it is enough to regularize the 
propagator \eqref{pr1} to take the situation under control. We use the following regularization
\begin{equation}\label{pr2}
\langle\,\xi^i(\varphi_1)\xi^j(\varphi_2)\rangle (\text{F})
=\frac{1}{2\pi}\Bigl(\gamma^{ij}
\sum_{n=1}^{\infty}\frac{\ep^{-\epsilon n}}{n}\ep^{in\varphi_{12}}
+\gamma^{ji}
\sum_{n=1}^{\infty}\frac{\ep^{-\epsilon n}}{n}\ep^{-in\varphi_{12}}\Bigr)
\quad.
\end{equation}
It is easy to see that in this case the sum in \eqref{bi} becomes a finite sum 
$I_0(\epsilon)=\sum_{n=1}^\infty \ep^{-\epsilon n}$. As a result, the partition function to this order is simply
\begin{equation}\label{pf3}
\text{Z}[\text{T,A},\epsilon]=\int [dx]\,\exp\bigl(-2\pi\text{T}+I_0(\epsilon)\tr\ln\gamma\bigr)
\quad.
\end{equation}
Taking into account the asymptotics $I_0(\epsilon)=\frac{1}{\epsilon}-\oh+O(\epsilon)$ \footnote{Some 
useful formulae can be found in appendix of \cite{AT}.}, one uses the minimal 
subtraction \footnote{ $(\mathbf {T},\mathbf {A})$ means the renormalized fields. We set the 
renormalized value of the tachyon field equal to zero, here and below.}
\begin{equation}\label{ms}
\text{T}=\frac{1}{2\pi\epsilon}\tr\ln\boldsymbol{\gamma}
\quad,\quad
\text{A}_i={\bf A}_i
\end{equation}
to get the desired BI factor in the expression for the renormalized partition function
\begin{equation}\label{pf4}
\text{\bf Z}[\mathbf{A}]=\int [dx]\,
\sqrt{\det\bigl(1+\mathbf{F}\bigr)}
\quad.
\end{equation}

To complete the story, let us look more closely at what happens within other regularizations. To get some insight, one 
can simply rescale $\epsilon$ as $\epsilon\rightarrow\epsilon (1+\epsilon)$. This modifies the asymptotics of 
$I_0$ as $I_0(\epsilon)=\frac{1}{\epsilon}-\frac{3}{2}+O(\epsilon)$. As a consequence, the minimal subtraction 
\eqref{ms} no longer results in the BI factor in {\bf Z}. The latter is of course unsurprising because the use of 
two different schemes can in general lead to different results. 

Let us now show how our proposal helps to fix the problem. We have for the  partition function \eqref{pf3} 
\begin{equation}\label{pf5}
\text{Z}[\text{T,A},\epsilon]=\int [dx]\,\exp\Bigl(-2\pi\text{T}
+\bigl[\frac{1}{\epsilon}-\frac{3}{2}+O(\epsilon)\bigr]
\tr\ln\gamma\Bigr)
\quad.
\end{equation}
In general, one can use arbitrary non-minimal subtractions like 
\begin{equation}\label{ms3}
\text{T}=\Bigl(\frac{1}{2\pi\epsilon}+c\Bigr)\tr\ln\boldsymbol{\gamma}
\quad,\quad
\text{A}_i=\mathbf{A}_i
\quad,
\end{equation}
where $c$ is constant. In this situation, the renormalized partition function $\text{\bf Z}$ is given by 
\begin{equation}\label{pf6}
\text{\bf Z}[\mathbf{A}]=\int [dx]\,\exp\Bigl(-(c+\frac{3}{2})
\tr\ln\boldsymbol{\gamma}
\Bigr)\quad.
\end{equation}
Noting that near the perturbative vacuum $S_{\text{eff}}\sim\text{\bf Z}$ and taking the 
following structures from $S_{\text{S-mat.}}$  
\begin{equation*}
\frac{1}{32}\mathbf{F}_{ij}^2\mathbf{F}_{nm}^2-
\frac{1}{8}\mathbf{F}_{ij}\mathbf{F}_{jn}\mathbf{F}_{nm}\mathbf{F}_{mi}
\end{equation*}
as the renormalization conditions, we find a unique solution for $c$, $c=-1$. As a result, 
we immediately get the BI factor 
in $\text{\bf Z}$ . Of course, it is well-known that the BI action reproduces the four point amplitudes of the gauge field 
but the novelty now is that we used the $\mathbf{F}^4$ terms to get the term in the effective action that includes all powers of the field strength.


We are now ready to carry out the analysis of the leading derivative corrections to 
the partition function, i.e. $Z_2$ term in Eq.\eqref{pf2}. The relevant contributions come from the 
following Feynman diagrams
%
\vspace{0.2cm}
\begin{figure}[ht]
\begin{center}
\includegraphics{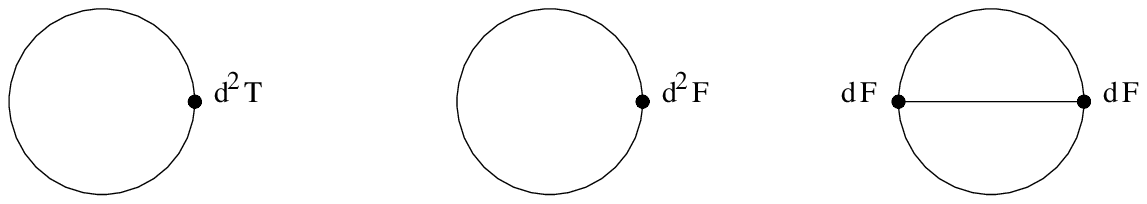}
\caption{Feynman diagrams that contribute the leading derivative corrections to the partition function.}
\label{fig:graph3}
\end{center}
\end{figure}
\vspace{-0.3cm} 

So we get to this order
\begin{equation}\label{pf7}
\begin{split}
\text{Z}[\text{T,A},\epsilon]&=\int [dx]\,\exp\bigl(-2\pi\text{T}+I_0(\epsilon)\tr\ln\gamma\bigr)
\Bigl[1-I_1(\epsilon)\,\pd_{ij}\text{T}\,\gamma^{ij}
+\frac{1}{2\pi}I_2(\epsilon)\,\pd_{nm}\text{F}_{ij}\,\gamma^{nm}\gamma^{ij}\\
&\quad
-\frac{1}{2\pi}I_2(2\epsilon)\,\pd_n\text{F}_{ij}\pd_m\text{F}_{kl}\,\gamma^{nm}\gamma^{ki}\gamma^{jl}
+\frac{1}{4\pi}I_3(2\epsilon)\,\pd_n\text{F}_{ij}\pd_m\text{F}_{kl}\,\gamma^{nm}\gamma^{ki}\gamma^{lj}
+O(\ap{}^2)
\Bigr]
\quad,
\end{split}
\end{equation}
where $I_1(\epsilon)=\sum_{n=1}^\infty \frac{\ep^{-\epsilon n}}{n}$, 
$I_2(\epsilon)=I_0 (\epsilon)I_1(\epsilon)$ and $I_3(\epsilon)=I_0(\epsilon)-I_1(\epsilon)$.

The renormalization at the leading order follows from our renormalization conditions already discussed  
above \footnote{We denote by $(\tilde{\text{T}},\tilde{\text{A}})$ the leading order renormalized fields.}
 \begin{equation}\label{ms4}
\text{T}=\tilde{\text{T}}+\frac{1}{2\pi}\bigl(I_0(\epsilon)+\oh\bigr)\tr\ln\tilde\gamma
\quad,\quad
\text{A}_i=\tilde{\text{A}}_i
\quad.
\end{equation}
We include non-singular terms to slightly simplify further calculations. Up to $O(\epsilon)$, the partition function 
then becomes
\begin{equation}\label{pf8}
\begin{split}
\text{Z}[\tilde{\text{T}},\tilde{\text{A}},\epsilon]&=\int [dx]\,\exp\bigl(-2\pi\tilde{\text{T}}
-\oh\tr\ln\tilde\gamma\bigr)
\Bigl[1
+\ln\epsilon\,\pd_{ij}\tilde{\text{T}}\,\tilde\gamma^{ij}
+\frac{1}{4\pi}\ln\epsilon\,\pd_{nm}\tilde{\text{F}}_{ij}\,\tilde\gamma^{nm}\tilde\gamma^{ij}\\
&\quad
-\frac{1}{4\pi}
\Bigl(\frac{\ln\epsilon}{\epsilon}-\frac{\ln2}{\epsilon}+\ln\epsilon+\ln2\Bigr)
\mathfrak{f}_3
+\frac{1}{4\pi}
\Bigl(\frac{1}{2\epsilon}+\ln\epsilon+\ln2-\oh\Bigr)
\mathfrak{f}_4
+O(\ap{}^2)
\Bigr]
\quad,
\end{split}
\end{equation}
with $\mathfrak{f}_3$ and $\mathfrak{f}_4$ as defined in \eqref{b}.

For future reference, it will help to know some general facts about 
$\pd{\text{F}}\pd{\text{F}}\gamma\gamma\gamma$ structures. So we pause here to 
shed some light on this issue. Using the Bianchi identity , it is possible to show that there are only seven independent 
structures among them
\begin{alignat}{3}
\mathfrak{f}_1&=\pd_n\text{F}_{ij}\pd_m\text{F}_{kl}\,\gamma^{nm}\gamma^{ij}\gamma^{kl}
\,\,\,\,,&\,\,\,\,
\mathfrak{f}_2&=\pd_n\text{F}_{ij}\pd_m\text{F}_{kl}\,\gamma^{nm}\gamma^{ik}\gamma^{jl}
\,\,\,\,,&\,\,\,\,
\mathfrak{f}_3&=\pd_n\text{F}_{ij}\pd_m\text{F}_{kl}\,\gamma^{nm}\gamma^{ki}\gamma^{jl}
\,\,\,\,\,\,, 
\notag\\
\mathfrak{f}_4&=\pd_n\text{F}_{ij}\pd_m\text{F}_{kl}\,\gamma^{nm}\gamma^{ki}\gamma^{lj}
\,\,\,\,, &\,\,\,\,
\mathfrak{f}_5&=\pd_n\text{F}_{ij}\pd_m\text{F}_{kl}\,
G^{ni}\theta^{kl}\theta^{jm}
\,\,\,\,,&\,\,\,\,
\mathfrak{f}_6&=\pd_n\text{F}_{ij}\pd_m\text{F}_{kl}\,
G^{ni}\theta^{kl}G^{jm}
\,\,\,\,, \notag
\\
\mathfrak{f}_7&=\pd_n\text{F}_{ij}\pd_m\text{F}_{kl}\,G^{ni}\Bigl(\gamma^{jl}\gamma^{mk}
+&\gamma^{lj}\theta^{mk}\Bigr)
\,\,\,\,,&
\phantom{AAA}&
\phantom{AAA}&
\label{b}
\end{alignat}
where $G$ and $\theta$ denote the symmetric and antisymmetric part of $\gamma$.  As we set the renormalized 
value of the tachyon field to zero, it allows us to transform all structures 
including a double derivative into the structures just described by the integration by parts. Moreover, 
a further simplification comes from the identity 
\begin{equation}\label{id}
0=\int [dx]\,\exp \bigl(-\oh\tr\ln\gamma\bigr)\pd_{ij}\theta^{ij}
\end{equation}
that reduces the total number of independent structures by one. To be more exact, $\mathfrak{f}_6$ drops 
from the integrand. 

Finally, setting the renormalization conditions to this order as \footnote{We follow the result for 
$S_{\text{S-mat.}}$ found in Ref.\cite{AT}.} 
\begin{equation}\label{rc}
-\frac{1}{48\pi}\pd_n\text{F}_{ij}\pd_n\text{F}_{ij}\text{F}_{kl}^2\,
-\frac{1}{6\pi}\pd_n\text{F}_{ij}\pd_n\text{F}_{jk}\text{F}_{kl}\text{F}_{li}\,
+\frac{1}{12\pi}\pd_n\text{F}_{ij}\pd_m\text{F}_{ij}\text{F}_{nk}\text{F}_{mk}
\end{equation}
we find a most general prescription for the renormalization
\begin{equation}\label{ms5}
\tilde{\text{T}}=\mathbf{T}+
\frac{1}{2\pi}\ln\epsilon\,\pd_{ij}\mathbf{T}\boldsymbol{\gamma}^{ij}+
\sum_{a=1, a\not=6}^7 b_a\mathfrak{f}_a(\mathbf{A})
\quad, \quad
\tilde{\text{A}}_i=\mathbf{A}_i-\Bigl(\frac{1}{2\pi}\ln\epsilon+2c_4\Bigr)\,
\pd_n\mathbf{F}_{ij}\mathbf{G}^{nj}
\quad,
\end{equation}
with
\begin{alignat*}{3}
b_1&=0
\quad&,\quad
b_2&=-\frac{1}{96\pi^2}
\quad&,\quad
b_3&=-\frac{1}{8\pi^2}\Bigl(\frac{\ln\epsilon}{\epsilon}-\frac{\ln2}{\epsilon}+\ln2-\oh\Bigr)
\quad,\\
b_5&=-\frac{1}{2\pi}(c_4+c_5)
\quad&,\quad
b_7&=-\frac{1}{2\pi}c_7
\quad&,\quad
b_4&=\frac{1}{8\pi^2}\Bigl(\frac{1}{2\epsilon}+\ln2-\oh\Bigr)-\frac{1}{2\pi}c_4
\qquad ,
\end{alignat*}
where $c_a$ are arbitrary numbers.

So we have for the partition function \footnote{Note that we set $\mathbf{T}=0$.}
\begin{equation}\label{pf9}
\begin{split}
\text{\bf Z}[\mathbf{A}]&=\int [dx]\,\sqrt{\det\bigl(1+\mathbf{F}\bigr)}
\Bigl[1+
\frac{1}{48\pi}\,
\pd_n\mathbf{F}_{ij}\pd_m\mathbf{F}_{kl}\,\boldsymbol{\gamma}^{nm}
\boldsymbol{\gamma}^{ik}\boldsymbol{\gamma}^{jl}
-\frac{1}{8\pi}\,
\pd_n\mathbf{F}_{ij}\pd_m\mathbf{F}_{kl}\,
\boldsymbol{\gamma}^{nm}\boldsymbol{\gamma}^{ki}\boldsymbol{\gamma}^{jl}
\\
&\quad
+c_5\,
\pd_n\mathbf{F}_{ij}\pd_m\mathbf{F}_{kl}\,\mathbf{G}^{ni}\boldsymbol{\theta}^{kl}
\boldsymbol{\theta}^{jm}
+c_7\,
\pd_n\mathbf{F}_{ij}\pd_m\mathbf{F}_{kl}\,\mathbf{G}^{ni}\Bigl(
\boldsymbol{\gamma}^{jl}\boldsymbol{\gamma}^{mk}
+\boldsymbol{\gamma}^{lj}\boldsymbol{\theta}^{mk}\Bigr)
\Bigr]
\,\,.
\end{split}
\end{equation}
To simplify the final expression, we have included the redefinition of the gauge field in \eqref{ms5}. It 
allows us to get rid of one structure, in the case of interest it is $\mathfrak{f}_4$.

To complete the story, let us write down the effective action. For example, the proposal \cite{W} is to take 
\begin{equation}\label{off}
S_{\text{eff}}=\bigl(1+\beta_i\frac{\pd}{\pd\mathbf{x}^i}\bigr)\text{\bf Z}(\mathbf{x}^i)
\quad.
\end{equation}
Here $\mathbf{x}^i$ mean the renormalized couplings of boundary interactions, i.e. the tachyon, gauge and other fields, 
$\beta_i$ are their RG $\beta$-functions and {\bf Z} is a renormalized partition function on the disk. However, 
in the problem of interest ($\mathbf{T}=0$, $\pd^2$-order) it exactly coincides with the off-shell partition 
function $\text{\bf Z}[\mathbf{A}]$, so it makes no sense to write it down again. Our statement is based on two facts: 
renormalizability \footnote{At this point we 
mean only logarithmic divergencies.} and dimensionless coupling. To be more precise, the renormalization of the 
gauge field is given by $\text{A}_i=\mathbf{A}_i+\delta\mathbf{A}_i$ with $\delta\mathbf{A}
\sim\ln\epsilon\,\beta_{\mathbf{A}}$, 
$\beta_{\mathbf{A}}\sim \mathbf{G}\pd\mathbf{F}$. A simple shift $\ln\epsilon\rightarrow\ln\epsilon +const$ or 
, equivalently the corresponding redefinition of the gauge field, results in
the second term in \eqref{off}. It is clear that this fails if one turns on massive fields or switches to next order in $\pd$. 

To conclude, let us make some remarks.

(i) Instead of doing the perturbation theory based on the propagator \eqref{pr1}, one can try to compute the 
partition function with the propagator \eqref{pr} and then make a resummation to bring the result into more 
simple form (see, e.g., \cite{AT}). It is clear that this way leads to much more technical difficulties than we had, so we do 
not study it in this paper. Let us only note that it leads to another expression for the effective action. Indeed, in the 
problem of interest the number of independent 
$\pd\text{F}\pd\text{F}\text{F}\text{F}$ structures turns out to be bigger by one than the number of independent 
$\pd\text{F}\pd\text{F}\gamma\gamma\gamma$ structures. Fortunately, the numbers of structures 
which are relevant for Eq.\eqref{i} are equal. It would be interesting to study whether this holds in higher orders too.
   
(ii) The use of the $\zeta$-function regularization simplifies the computation but it disguises the origin of the 
problem as there are no power like singularities in this regularization. For example, the expression for the 
partition function \eqref{pf7} now translates to 
\begin{equation}\label{pfz}
\begin{split}
\text{Z}[\text{T,A},\epsilon]&=\int [dx]\,\exp\bigl(-2\pi\text{T}\bigr)
\sqrt{\det\bigl(1+\text{F}\bigr)}
\Bigl[1-\zeta(1)\,\pd_{ij}\text{T}\,\gamma^{ij}
-\frac{1}{4\pi}\zeta(1)\,\pd_{nm}\text{F}_{ij}\,\gamma^{nm}\gamma^{ij}\\
&\quad
+\frac{1}{4\pi}\zeta(1)\,\pd_n\text{F}_{ij}\pd_m\text{F}_{kl}\,\gamma^{nm}\gamma^{ki}\gamma^{jl}
-\frac{1}{4\pi}\bigl(\zeta(1)+\oh\bigr)\,
\pd_n\text{F}_{ij}\pd_m\text{F}_{kl}\,\gamma^{nm}\gamma^{ki}\gamma^{lj}
+O(\ap{}^2)
\Bigr]
\quad,
\end{split}
\end{equation}
where $\zeta(z)=\sum_{n=1}\frac{1}{n^z}$ and $\zeta(0)=-\oh$.

Since there is no need in the renormalization of the power divergence as in \eqref{ms4}, the use of 
the minimal subtraction seems rather natural at the leading order. However, such a 
scheme ($\zeta$-regularization with 
minimal subtraction) fails at next order. This can be fixed by using a non-minimal subtraction like \eqref{ms5} 
which is consistent with our renormalization conditions \eqref{rc}. As a result, we again end up with the 
expression \eqref{pf9}.

(iii) Finally, let us note that the question on the renormalization conditions was also raised by Li and Witten 
within the background independent open string theory \cite{W} whose setting is rather similar to what we 
discussed above. We do not have a final solution of this problem to offer today. What we proposed can be only 
used to fix a part of the partition function. 


{\it Acknowledgements.} 
We would like to thank H. Dorn and A. A. Tseytlin for useful discussions and reading the manuscript. This 
work is supported in part by DFG under Grant No. DO 447/3-1 and the European Commission RTN Programme 
HPRN-CT-2000-00131.



\end{document}